\documentstyle[preprint,prl,aps]{revtex}

\newcommand{\be}{\begin{equation}}
\newcommand{\ee}{\end{equation}}
\newcommand{\ben}{\begin{eqnarray}}
\newcommand{\een}{\end{eqnarray}}

\newcommand{\nd}{\noindent}

\begin{document}
\draft

\title{Jensen-Shannon divergence, Fisher information, and Wootters' hypothesis}

\author{M. Casas$^{1}$, P. W. Lamberti$^{2,\,4}$, A. Plastino$^{3,\,4}$,
and A. R. Plastino$^{3,\,4,\,5}$}

\address{$^1$ Department de Fisica and IMEDEA, Universitat de les Illes Balears
\\
07122 Palma de Mallorca, Spain \\ $^2$ Facultad de Matem\'atica,
Astronom\'{\i}a y F\'{\i}sica - Universidad Nacional de C\'ordoba
\\Ciudad Universitaria - 5000 C\'ordoba, Argentina \\$^3$ Universidad Nacional d
e La
Plata, C.C. 727, 1900 La Plata, Argentina \\
$^4$ Argentine National Research Center (CONICET), C. C. 727, 1900
La Plata, Argentina\\ $^5$Department of Physics, University of
Pretoria, Pretoria 0002, South Africa}

\maketitle

\begin{abstract}

We discuss different statistical distances in probability space,
with emphasis on the Jensen-Shannon divergence, vis-a-vis {\it
metrics} in Hilbert space and their relationship with Fisher's
information measure. This study provides further reconfirmation of
Wootters' hypothesis concerning the possibility that statistical
fluctuations in the outcomes of measurements be regarded (at least
partly) as responsible for the Hilbert-space structure of quantum
mechanics.



\end{abstract}
\maketitle

\section{Introduction}

Wootters \cite{wootters} has shown that nature defines the
distance $D_W$ between two quantum states, say, $\psi_1$ and
$\psi_2$, by counting the number of distinguishable {\it
intermediate} states, which establishes  {\it a link} between {\it
statistics} (distances in probability spaces, determined by the
size of statistical fluctuations \cite{wootters}) and {\it
geometry} (metrics in Hilbert space) that has been considerably
strengthened in \cite{caves}. The claim is made, however, that a
proper understanding of this link needs still further elaboration
\cite{wootters}, raising the interesting possibility that
statistical fluctuations in the outcomes of measurements might be
partly responsible for the Hilbert-space structure of quantum
mechanics. An illuminating expansion of the work of
\cite{wootters} is provided in \cite{caves} and some footnotes can
be found in  \cite{ravi}. Quite lucid, insightful,  and also
didactic
  mathematical treatments pertaining to the field of information geometry
   are those of
   Refs. \cite{amari,gill,murray,brody} (see also references therein).

Obviously, the concept of distance between different rays in (the
same) Hilbert space plays an important role in a variety of
circumstances, like different preparations of the same system
\cite{wootters} or the geometric properties of the quantum
evolution sub-manifold \cite{abe}. It is also relevant, for
instance, in discussing squeezed coherent states, displaced number
states, generalized coherent spin states, etc. \cite{abe}, or for
ascertaining the quality of approximate treatments \cite{ravi}.
One encounters it as well in connection with the detection of weak
signals.

Inspired by the pioneer effort of Wootters \cite{wootters}, a set
of well-known Hilbert space metrics were compared, in Ref.
\cite{ravi}, to that underlying Wootters' $D_W$. Let a quantum
state $\psi_{\alpha}({\bf x})$ be parameterized by $n$ real
parameters collectively denoted by the symbol $\alpha$. A set of
rays corresponding to the states with all possible $\alpha$-values
constitutes an $n$-dimensional manifold ${\cal K}$ of the Hilbert
space ${\cal H}$. Set $\psi_1=\psi_{\alpha}({\bf x})$ and
$\psi_2=\psi_{\alpha+\Delta\alpha}({\bf x})$. Expanding now
$\psi_2$ in a $\Delta\alpha$-series, it was found in \cite{ravi}
that, up to second order in $\Delta\alpha$, several Hilbert space
distances between $\psi_2$ and $\psi_1$  coincide with Wootters'
$D_W$, as does also the Kullback's cross-entropy \cite{Kullback51}
between $\vert \psi_2\vert^2$ and $\vert \psi_1\vert^2$.

Here we will pursue the work of \cite{ravi}  i) by incorporating
additional measures like the {\it Jensen-Shannon divergence} and,
more importantly,  ii) by providing still another interpretation
to the role of fluctuations on which the work of Wootters'
originally focused attention.


 \section{Fisher's Information measure}

An information measure can  primarily be viewed as a a quantity
that characterizes a given probability distribution (PD) $\vec P$
\cite{Shannon48,katz}. Shannon's logarithmic information measure
\cite{Shannon48}
\begin{equation}
S[\vec P]~=~-\sum_{j=1}^N~p_j~\ln(~p_j~) \ , \label{ein}
\end{equation}
is regarded as the  measure of the uncertainty associated to
probabilistic physical processes described by the probability
distribution $\{ p_j,~j=1, \cdots , N \}$ (${\vec P} \equiv ( p_1,
p_2,\cdots, p_N)$ the probability vector in the probability space
$\Omega \subset{\cal R}^{N}$). We will be concerned here mainly
with another important information-theoretic
 measure: that of Fisher's ($I$) \cite{Frieden,roybook},  advanced by  R.~A.
  Fisher in the twenties.

  $I$ has been the subject of much work lately (a detailed study can be found in
references~\cite{Frieden,roybook}).
   Let us consider a system that is specified by a physical
  parameter $\alpha$,  while {\bf x} is a stochastic variable 
$({\bf x}\,\in\,\Re^{N})$
  and
  $P_\alpha({\bf x})$ the probability density for ${\bf x}$,
  which depends on the parameter $\alpha$.  An observer makes a
  measurement of
   ${\bf x}$ and
  has to best infer $\alpha$ from this  measurement,
   calling the
    resulting estimate $\tilde \alpha=\tilde \alpha({\bf x})$. One
   wonders how well $\alpha$ can be determined. Estimation theory~\cite{cramer}
   asserts that the best possible estimator $\tilde
   \alpha({\bf x})$, after a very large number of ${\bf x}$-samples
  is examined, suffers a mean-square error $e^2$ {\bf for} $\alpha$ that
  obeys a relationship involving Fisher's $I$, namely, $Ie^2=1$,
  where the Fisher information measure $I$ is of the form
  \be
  I(\alpha)=\int \,\mathrm{d}{\bf x}\,P_\alpha({\bf
  x})\left\{\frac{\partial \ln P_\alpha({\bf x})}{
  \partial \alpha}\right\}^2  \label{ifisher}.
  \ee

  This ``best'' estimator is called the {\it efficient} estimator.
  Any other estimator must have a larger mean-square error. The only
  proviso to the above result is that all estimators be unbiased,
  i.e., satisfy $ \langle \tilde \alpha({\bf x}) \rangle=\,\alpha
  \label{unbias}$.   Thus, the inverse of Fisher's
  information measure provides a lower bound for the
  mean-square error $e^2$ associated with the
   statistical inference of the parameter $\alpha$. No matter what
   estimator we use (as long as it is an unbiased estimator)
  we have $e^2\,\ge \,1/I.$ This inequality is referred to
  as the Cramer-Rao bound \cite{roybook}.

  \section{The Jensen-Shannon divergence}

  We review now some other measures with which we will be concerned in the prese
nt
  work and were not taken into account in \cite{ravi}.
    Let ${\vec P}_{(k)} \in  \Omega \subset{\cal R}^{N}$, with $k=1,2$, denote
two different probability distributions for a particular set of
$N$ accessible states.  The components of the two probability
vectors ${\vec P}_{(k)}$ must satisfy the following two
constraints: {\it a)\/} $\sum_{j=1}^N~p_j^{(k)}=1$ and {\it b)\/}
$0 \leq p_j^{(k)} \leq 1~\forall j$. The set $\Omega$ defined by
these constraints is the simplex $S_N$, which is a convex
$(N-1)$-dimensional subset of $R^N$. A quite important,
information-theoretical based divergence measure between ${\vec
P}_{(1)}$ and ${\vec P}_{(2)}$ was originally introduced by Lin
\cite{lin} that  came afterwards to be called the Jensen-Shannon
divergence (JSD) \cite{pregiven,lamberti} that
\begin{itemize} \item induces a {\it true} metric in $\Omega
\subset{\cal R}^{N}$, being indeed the square of a metric
\cite{topsoe}, and \item is intimately related to the
Kullback-Leibler relative entropy  $K$  for two probability
distributions ${\vec P}_{(1)}$ and ${\vec P}_{(2)}$, given by
\cite{Kullback51}
\begin{equation}
K[{\vec P}_{(1)} \vert {\vec P}_{(2)}]~= ~\sum_j~p^{(1)}_j~\ln
\left(~p^{(1)}_j~/~p^{(2)}_j~\right). \label{KL-entropy}
\end{equation}
\end{itemize}
We first define
\begin{equation}
\label{uno1} J_0\left[ {\vec P}_{(1)}, {\vec P}_{(2)} \right]~=~
K\left[ {\vec P}_{(1)} \Biggm| \left( {\frac{1}{2}}{\vec
P}_{(1)}~+~
                                       {\frac{1}{2}}{\vec P}_{(2)}
                               \right) \right] \ ,
\end{equation}
and then the symmetric quantity
\begin{eqnarray}
\label{dos2} J_1\left[{\vec P}_{(1)}, {\vec P}_{(2)} \right]&=&
J_0\left[{\vec P}_{(1)}, {\vec P}_{(2)} \right] + J_0\left[{\vec
P}_{(2)}, {\vec P}_{(1)} \right]  \\ \nonumber &=& 2~S\left[
\frac{1}{2} {\vec P}_{(1)}~+~
              \frac{1}{2} {\vec P}_{(2)} \right]
   - S\left[ {\vec P}_{(1)} \right]
   - S\left[ {\vec P}_{(2)} \right].
\end{eqnarray}

Let now $\pi_1,~\pi_2 > 0$; $\pi_1+\pi_2=1$ be the ``weights" of,
respectively, the probability distributions ${\vec P}_{(1)},~{\vec
P}_{(2)}$. The JSD reads
\begin{equation}
\label{tres3} J^{\pi_1,\pi_2}\left[{\vec P}_{(1)}, {\vec P}_{(2)}
\right]~=~ S\left[ \pi_1 {\vec P}_{(1)}~+~
        \pi_2 {\vec P}_{(2)} \right]
   - \pi_1~S\left[ {\vec P}_{(1)} \right]
   - \pi_2~S\left[ {\vec P}_{(2)} \right]
\end{equation}
 which is a positive-definite quantity that vanishes iff ${\vec
P}_{(1)} = {\vec P}_{(2)}$ almost everywhere
\cite{pregiven,lamberti}. In the particular case $\pi_1=\pi_2= 1/2
$ the measure (\ref{tres3}) is symmetric. Notice also that
$J^{\frac{1}{2},\frac{1}{2}}=\frac{1}{2}J_1$.

Using the q-information measures it is possible to construct a 
q-Kullback-Leibler relative entropy for two probability distributions.
In particular, from the Tsallis entropy the q-Kullback entropy reads
\begin{equation}
K^T_q[{\vec P}_{(1)} \vert {\vec P}_{(2)}]~= ~-\sum_j~p^{(1)}_j~\ln
\left(~p^{(1)}_j~/~p^{(2)}_j~\right). \label{q-KL-entropy}
\end{equation}

\section{Present results}

Suppose that a quantum state $\psi(\bf x)$ is parameterized, as
stated in the Introduction,  by a real parameter that we denote by
$\alpha$, and write for simplicity $\psi_{\alpha}(\bf x) \equiv
\psi(\alpha)$. We consider two states that differ by a change
$\alpha \to \alpha + \Delta \alpha$ and expand the second one up
to second order in $\Delta \alpha$ [we set $\Delta ^2 \alpha =
(\Delta \alpha)^2]$.

\begin{equation}
\label{wave} \vert \psi(\alpha+ d\Delta \alpha)> = \vert
\psi(\alpha)> +\Delta \alpha \frac {d}{d\alpha}\vert \psi(\alpha)>
+ \frac {1}{2}\Delta^2 \alpha\frac {d^2}{d\alpha^2}\vert
\psi(\alpha)>+ ...
\end{equation}

\nd We wish to compare these associated wave functions (say,
$\psi_1$ and $\psi_2$), by recourse to different measures. To this
end, and following Eq. (\ref{wave}), we expand $\psi(\alpha +
\Delta \alpha)$ up to second order and assume that both
$\psi(\alpha)$ and $\psi(\alpha + \Delta \alpha)$ are properly
normalized to unity. For the present purposes we set $P_{(1)} =
\vert\psi(\alpha)\vert^2$ and $P_{(2)} = \vert\psi(\alpha + \Delta
\alpha)\vert^2$. The symmetrized Kullback-Leibler relative entropy
\begin{equation}\label{kullbacks}
K_S = K[\vec{ P}_{(1)}\vert\vec{ P}_{(2)}] + K[\vec{ P}_{(2)}\vert
\vec{ P}_{(1)}]
\end{equation}

\nd is given by

\begin{eqnarray} \label{kullbacks1}
K_S &=& \int dx \vert \psi(\alpha) \vert^2 \ln \left[\frac{\vert
\psi(\alpha) \vert^2}{\vert \psi(\alpha + \Delta \alpha)
\vert^2}\right]+ \nonumber \\
&& \int dx \vert \psi(\alpha + \Delta \alpha) \vert^2 \ln
\left[\frac{\vert \psi(\alpha + \Delta \alpha) \vert^2}{\vert
\psi(\alpha ) \vert^2}\right]
\end {eqnarray}

\nd For simplicity's sake we restrict ourselves to one dimensional
problems (the essentials of our discourse can already be
apprehended at this stage) where, for stationary cases, one always
can assume, without loss of generality, that wave functions are
real. Up to second order in $\Delta^2\alpha$ ($\psi' =
\partial \psi /\partial \alpha$)
\begin{equation}\label{order2}
[\psi(\alpha + \Delta \alpha)]^2 = \psi^2\left[1+ \Delta \alpha
\frac{\partial \ln \psi^2}{\partial\alpha}+ \Delta^2 \alpha
\left[\frac{\psi''}{\psi} +
\left(\frac{\psi'}{\psi}\right)^2\right] \right],
\end{equation}
so that

\begin{equation}\label{order22}
P_{(2)} = P_{(1)}\left[1+ \Delta \alpha  \frac{\partial \ln
P_{(1)}}{\partial \alpha}+\frac{\Delta^2
\alpha}{2}\frac{P''_{(1)}} {P_{(1)}}+ ...\right],
\end{equation}
and we recast the Kullback-Leibler measure in the fashion

\begin{eqnarray}\label{ks}
K_S &=& - \int dx P_{(1)} \ln \left[[1+ \Delta \alpha
\frac{\partial \ln P_{(1)}}{\partial \alpha}+\frac{\Delta^2
\alpha}{2}\frac{P''_{(1)}} {P_{(1)}}\right]\nonumber\\
 &&+\int dx
P_{(1)}\left[1+ \Delta \alpha  \frac{\partial \ln
P_{(1)}}{\partial \alpha}+\frac{\Delta^2
\alpha}{2}\frac{P''_{(1)}} {P_{(1)}}\right] \nonumber\\
&& \ln \left[1+ \Delta \alpha \frac{\partial \ln P_{(1)}}{\partial
\alpha}+\frac{\Delta^2 \alpha}{2}\frac{P''_{(1)}}
{P_{(1)}}\right].
\end{eqnarray}
 We will appeal to an expansion of  $\ln(1+y)$ up to first order in $y$. 
Remember  that
both $ \psi(\alpha)$  and $\psi(\alpha+ \Delta \alpha)$ are
properly normalized, which implies
\begin{equation}\label{nul1}
\Delta \alpha \int dx P_{(1)} \frac{\partial \ln P_{(1)}}{\partial
\alpha}=0,
\end{equation}
and
\begin{equation}\label{nul2}
(\Delta^2 \alpha/2) \int dx P''_{(1)}=0,
\end{equation}
so that one obtains
\begin{equation}\label{Ks2}
K_S= \Delta^2 \alpha \int dx P_{(1)} \left(\frac{\partial \ln
P_{(1)}}{\partial \alpha}\right)^2 = \Delta^2 \alpha I (\alpha),
\end{equation}
where $I(\alpha)$ is the Fisher information measure defined in
(\ref{ifisher}). This relation is to be expected on
information-theoretic grounds \cite{roybook}, but, within the
present context, it was not discussed neither in \cite{ravi} nor
in \cite{wootters}, so we wish to place strong emphasis on
(\ref{Ks2}). Notice also that the Kullback Leibler measure is
stable against first order changes in $\Delta \alpha$ \cite{ravi}.

\nd  Let us tackle now a subject that is new for the present
scenario:
  the Jensen-Shannon divergence (\ref{uno1}). Up to the second order in $\Delta
  \alpha$ one has: i) for $J_0$
\begin{equation}\label{j02}
J_0\left[{ P}_{(1)}, {P}_{(2)}\right] = \frac{1}{8}\Delta^2 \alpha
\int dx P_{(1)} \left(\frac{\partial \ln P_{(1)}}{\partial
\alpha}\right)^2 = \frac{1}{8}\Delta^2 \alpha I (\alpha),
\end{equation}
ii) for  the symmetric $J_1$ of (\ref{dos2})
\begin{equation}\label{j12}
J_1\left[{P}_{(1)}, {P}_{(2)}\right] = \frac{1}{4}\Delta^2 \alpha
\int dx P_{(1)} \left(\frac{\partial \ln P_{(1)}}{\partial
\alpha}\right)^2 = \frac{1}{4}\Delta^2 \alpha I (\alpha),
\end{equation}
and iii) for the quantity  defined in (\ref{tres3})
\begin{equation}\label{jsd2}
J^{\pi_1,\pi_2}\left[{ P}_{(1)}, { P}_{(2)} \right]= \frac{\pi_1
\pi_2}{2}\Delta^2 \alpha \int dx P_{(1)} \left(\frac{\partial \ln
P_{(1)}}{\partial \alpha}\right)^2 = \frac{\pi_1 \pi_2}{2}\Delta^2
\alpha I (\alpha),
\end{equation}
i.e.,  all the Jensen-Shannon divergences (\ref{j02}),
(\ref{j12}), and (\ref{jsd2}), together with  the Kullback-Leibler
measure (\ref{Ks2}), are proportional to $I$, {\it our main result
thusfar}.  Now, it was  shown in \cite{ravi} that the Euclidean
distance between neighboring states can be evaluated as
\begin{equation}\label{euclidean}
dS^2_E =\int dx\left[ \psi(\alpha+\Delta
\alpha)-\psi(\alpha)\right]^2=\Delta^2 \alpha \int dx \left\vert
\frac{\partial \psi}{\partial \alpha}\right\vert^2,
\end{equation}
so that it easily follows that, up to second order in $\Delta
\alpha$, this distance  is also proportional to $I$, that is well
known to be a  measure of the gradient of the
probability-amplitude \cite{roybook}
\begin{equation} \label{eu2}
dS^2_E =  \frac{1}{4} \Delta^2 \alpha I (\alpha),
\end{equation}
which  {\it in the present context} can be regarded as a new
result. Additionally, by recourse to the comparison between
$\psi(\alpha)$ and $\psi(\alpha+\Delta\alpha)$, we can relate the
Euclidean distance with the Wootters one  $D_W$. Using the
following result from \cite{ravi}
\begin{equation}\label{sew}
dS^2_E = 2(1-<\psi (\alpha) \vert \psi (\alpha + \Delta \alpha)>)
= 2(1 - \cos \gamma),\end{equation} where $\gamma$ is a small
angle, and now expanding $\cos  \gamma$ up to order $\gamma^2$ one
obtains

\begin{equation} \label{w2}
dS^2_E \cong \gamma^2 = [\arccos (<\psi (\alpha) \vert \psi
(\alpha + \Delta \alpha)>)]^2 = dS^2_W= D_W
\end{equation}
where $D_W\equiv dS^2_W$ is the Wootters distance \cite{ravi}, so
that Eq.(\ref{eu2}) is tantamount to
\begin{equation}\label{w3}
dS^2_W =\frac{1}{4} \Delta^2 \alpha I (\alpha),
\end{equation}
which can also regarded as a new result within the current
context. We pass now to the celebrated Fubini-Study metric
\cite{fubini,pati}. Considering  neighboring states we have
\cite{ravi}

\begin{equation}\label{f1}
dS^2_F = 1- \vert<\psi(\alpha)\vert \psi(\alpha+\Delta
\alpha)>\vert^2,
\end{equation}
which, after our by now familiar expansion up to second order
yields
\begin{equation}\label{f2}
dS^2_F=\Delta^2 \alpha \int dx \left\vert \frac{\partial
\psi}{\partial \alpha}\right\vert^2 = \frac{1}{4} \Delta^2 \alpha
I (\alpha).
\end{equation}
Again, the Euclidean distance, and also  the Wootters' and
Fubini-Study ones, are stable against first order changes in
$\Delta \alpha$ and all of them  coincide,  up order $\Delta^2
\alpha$, with the $J_1$ Jensen-Shannon divergence, {\it still
another new result}. All these distances  are proportional to the
concomitant Fisher measure.

\section{Discussion}
Let us examine a bit more closely the problem of estimating a
single
  parameter ($\alpha$) of a system or phenomenon from knowledge of some
  measurements of the variable $x$ \cite{roybook}. Consider that we have at our
  disposal $N$ data values of this variable $x_1,\ldots,x_N \equiv {\bf
  x}.$ The system or phenomenon is governed by the conditional probability law
  (likelihood law) $f({\bf x}\vert \alpha) \equiv f_\alpha({\bf
  x}).$ The data obey

  \be \label{data} {\bf x}=\alpha + {\bf y};\,\,\,\,({\bf y}\,\,{\rm added\,\,no
ise\,\,values},
 ), \ee with {\bf y} assumed to be intrinsic to the parameter
  $\alpha$ under measurement. As an example,  $\alpha$ could be
  a particle's  position and ${\bf y}$ its concomitant
  fluctuations. The system consisting of quantities $\alpha,\,{\bf y},\, {\bf
  x}$  is a closed one \cite{roybook}. The data are used in an
  estimation principle to form an estimate $\tilde\alpha$  of $\alpha$ which is 
a function of all the data
  (say,
  $\sum_i^N\,x_i/N$), and one assumes that the overall measurement
  procedure is ``smart" in the sense that $\tilde\alpha$ is on
  average a better estimate of $\alpha$ than any of the data observables
  \cite{roybook}. We see then that, on account of the Cramer-Rao bound $I$
  carries information with regards to intrinsic uncertainties,
  which, quantum mechanically, correspond to intrinsic
  fluctuations \cite{roybook}. It is {\it in this light that we
  have to regard the results of the preceding Section}.

\nd The metric structure of the manifold ${\cal K}$ is completely
expressed by the uncertainties and correlations of Hermitian
operators generating various evolutions of a given quantum state
\cite{abe} which are neatly captured by the Fubini metric, as has
been demonstrated by using squeezed states \cite{abe2}. But this
metric does coincide, up to second order, with all the others
considered here, and with $D_W$ in particular, that gave rise to
the Wootters' suggestion mentioned in the Introduction: {\it
statistical fluctuations in the outcomes of measurements might be
partly responsible for the Hilbert-space structure of quantum
mechanics}. This view is now considerably strengthened in
discovering that all distances (here considered) between quantum
neighboring states, whether of statistical or Hilbert's metric
origin, are proportional to Fisher's measure, up to second order
approximation. Now, \begin{itemize} \item since $I$ captures, as
pointed out above, the {\it essentially fluctuating nature} of the
variables ${\bf x}$ on which the state $\psi_{\alpha}({\bf x})$
depends, and \item distances between neighboring states are
proportional to $I$, it follows that \item Wootters' viewpoint
receives yet further (independent) reconfirmation.
\end{itemize}

{\bf Acknowledgments:} This work was partially supported by the 1)
AECI Scientific Cooperation Program, 2) MCyT grant BFM2002-03241 and
3)  Grant BIO2002-04014-C03-03 (Spanish Government). One of us
(PWL) acknowledges  financial support from  SECYT (Universidad
Nacional de Cordoba, Argentina).

\end{document}